\documentstyle[aps,epsf,epsfig]{revtex}
\newcommand{\be}{\begin{equation}}
\newcommand{\ee}{\end{equation}}
\newcommand{\bea}{\begin{eqnarray}}
\newcommand{\eea}{\end{eqnarray}}
\newcommand{\ba}{\begin{eqnarray}}
\newcommand{\ea}{\end{eqnarray}}
\newcommand{\nn}{\nonumber \\}
\newcommand{\eqn}[1]{(\ref{#1})}
\newcommand{\beq}{\begin{equation}}
\newcommand{\eeq}{\end{equation}}
\newcommand{\beqa}{\begin{eqnarray}}
\newcommand{\eeqa}{\end{eqnarray}}
\newcommand{\beqar}{\begin{eqnarray*}}
\newcommand{\eeqar}{\end{eqnarray*}}
\newcommand{\e}{\epsilon}
\newcommand{\reef}[1]{(\ref{#1})}

\newcommand{\ie}{{\it i.e.,}\ }

\newcommand{\rom}[1]{{\mathrm{#1}}}
\font\mybb=msbm10 at 10pt
\def\bb#1{\hbox{\mybb#1}}

\def\bR {\bb{R}}

\def\sac{\, , \qquad}

\begin{document}

\twocolumn[\hsize\textwidth\columnwidth\hsize\csname
@twocolumnfalse\endcsname

\rightline{NSF-KITP-04-84}
\rightline{hep-th/0407065}

\title{A Supersymmetric Black Ring}
\author{Henriette Elvang,$^1$ Roberto Emparan,$^2$
David Mateos,$^3$ and Harvey S. Reall$^4$ \\}
\address{$^1\,$Department of Physics, University of California, Santa
Barbara, CA 93106-9530, USA \\ 
$^2$\,ICREA,
Departament de F{\'\i}sica Fonamental, and \\
C.E.R. en Astrof\'{\i}sica, F\'{\i}sica de Part\'{\i}cules i Cosmologia,
Universitat de Barcelona, Diagonal 647, E-08028 Barcelona, Spain \\
$^3\,$Perimeter Institute for Theoretical Physics, Waterloo,
Ontario N2J 2W9, Canada \\
$^4\,$Kavli Institute for Theoretical Physics, University of
California, Santa Barbara, CA 93106-4030, USA \\
{\small elvang@physics.ucsb.edu, emparan@ub.edu,
dmateos@perimeterinstitute.ca, reall@kitp.ucsb.edu} \\
}
\maketitle
\begin{abstract}
A new supersymmetric black hole solution of five-dimensional
supergravity is presented. It has an event horizon of topology $S^1
\times S^2$. This is the first example of a supersymmetric,
asymptotically flat black hole of non-spherical topology. The solution
is uniquely specified by its electric charge and two independent angular
momenta. These conserved charges can be arbitrarily close,
but not exactly equal, to those of a supersymmetric black hole of
spherical topology.
\end{abstract}
\vskip1pc]


A major success of string theory is the statistical-mechanical
explanation of the Bekenstein-Hawking entropy of certain supersymmetric
black holes. 
The original example is the five-dimensional black hole studied in
\cite{stva}. This is also the simplest example, as it carries the
minimum number of net charges necessary to have a finite-area regular
horizon, namely D1- and D5-brane charges and linear momentum along an
internal direction.
A generalized solution with the same charges and equal angular momenta
in two orthogonal planes was discovered, and its entropy microscopically
reproduced, by Breckenridge, Myers, Peet and Vafa (BMPV) \cite{bmpv},
thus extending the success of \cite{stva} to rotating black holes with a
single independent rotation parameter.

The BMPV black hole has a topologically spherical event
horizon. It has recently been realized that this is not true
of all five-dimensional rotating black holes: the vacuum Einstein
equations admit a (non-supersymmetric) {\it black ring} solution, with
horizon topology $S^1\times S^2$ \cite{ER}. 
The existence of black rings raises the question of whether there
are any {\it supersymmetric} black holes in five dimensions other
than BMPV. 

In \cite{reall:02} it was proven that the geometry of the event horizon
of any supersymmetric black hole of minimal five-dimensional
supergravity must be (i) $T^3$, (ii) $S^1 \times S^2$ or (iii) (possibly
a quotient of) a homogeneously squashed $S^3$. It was also proven that
the only asymptotically flat supersymmetric solution with horizon
geometry (iii) is the BMPV black hole (which reduces to a
solution of minimal supergravity when its three charges are set equal).
The purpose of this letter is to present a solution of type (ii), that is, a {\it supersymmetric black ring}. Such a solution was conjectured to exist in \cite{benakraus} motivated by the work of \cite{mathur}.  

This is the first example of an asymptotically flat supersymmetric
solution with a regular event horizon of non-spherical topology. It
possesses a richer structure than the BMPV solution, which we will
see arises as a particular case. It is parametrized by its electric charge and
two {\it independent} angular momenta, which illustrates the fact
that supersymmetry imposes no constraint on the angular momenta.
It also has a non-vanishing magnetic dipole, which is fixed by the
asymptotic charges and therefore is not an independent parameter. 
Some black rings are believed to be unstable \cite{ER} but 
supersymmetry should ensure that this new solution is stable. 

Our solution corresponds to taking equal values for the three charges
(D1, D5 and momentum) and three dipoles (D1, D5 and Kaluza-Klein monopole)
of a more general supersymmetric black ring (or, viewed in higher 
dimensions, a black supertube \cite{MT} with three charges \cite{benakraus}).
The details of these will be given elsewhere \cite{EEMR}, but we do
anticipate that, although the equal-charge solution presented here is
entirely determined by its conserved charges, this is not the case for
those of \cite{EEMR}.

Progress in understanding how the string microscopic description of
black holes distinguishes between different horizon topologies has
recently been made \cite{HE}. The existence of the supersymmetric black
ring opens for the first time the exciting possibility of studying this
question for a black hole with a regular horizon of finite area in a
supersymmetric, highly controlled, setting. We leave this and other
questions raised by the existence of our solution for the future.


\noindent {\bf The solution.}
The bosonic sector of five-dimensional minimal supergravity is
Einstein-Maxwell theory with a Chern-Simons term.
Any supersymmetric solution of this theory
must possess a non-spacelike Killing vector field $V$
\cite{gibbons:94}. In a region where $V$ is time-like,
the metric can be written
as \cite{harveyetal}
\be
ds^2 = -f^2(dt+\omega)^2 + f^{-1} ds^2(M_4) \,, \label{metric}
\ee
where $V = \partial/\partial t$, $M_4$ is an arbitrary hyper-K\"ahler
space, and $f$ and $\omega$ are a scalar and a one-form on $M_4$,
respectively, which must satisfy \cite{harveyetal}
\be
d G^+ =0 \sac \Delta f^{-1} = \frac{4}{9} (G^+)^2 \,, \label{eom}
\ee
where $G^+\equiv \frac{1}{2}f(d\omega+\star d\omega)$, with $\star$
the Hodge dual on $M_4$ and $\Delta$ is the Laplacian on $M_4$.

For our solution, $M_4$ is just flat space $\bR^4$, whose metric we
write as \cite{ER,HE}
\bea
ds^2(\bR^4) = \frac{R^2}{(x-y)^2} &\Big[& \frac{dy^2}{y^2-1} +
(y^2-1)d\psi^2 \nn
&& +\frac{dx^2}{1-x^2}+(1-x^2)d\phi^2 \Big] \,.
\label{base}
\eea
The coordinates have ranges $-1\leq x\leq 1$ and $-\infty<y\leq -1$,
and $\phi,\psi$ have period $2\pi$. 
Asymptotic infinity lies at $x\to y\to -1$. Note that the apparent
singularities at $y=-1$ and $x = \pm 1$ are merely coordinate
singularities, and that $(x,\phi)$ parametrize (topologically) a
2-sphere. The locus $y=-\infty$ is, in \reef{base}, a circle of radius
$R>0$ parametrized by $\psi$. In the full geometry \reef{metric} it will
be blown up into a ring-shaped horizon. The orientation is
$\epsilon_{y\psi x \phi} \equiv 1$. 

The scalar and one-form of the solution are given by
\be
f^{-1}=1+\frac{Q-q^2}{2R^2}(x-y)-\frac{q^2}{4 R^2}(x^2-y^2)
\label{simplef}
\ee
and $\omega= \omega_{\psi}(x,y) d\psi+\omega_{\phi}(x,y) d\phi$,
with
\beqa
\label{omegas}
\omega_\phi &=& -\frac{q}{8R^2} (1-x^2) \left[3Q - q^2 ( 3+x+y)
 \right]\,, \\
\omega_\psi &=& \frac{3}{2} q(1+y)  + \frac{q}{8R^2} (1-y^2)
\left[3Q - q^2 (3 +x+y) \right]
\,.\nonumber
\eeqa
$Q$ and $q$ are positive constants, proportional to the net charge and
to the local dipole charge of the ring, respectively. We assume
$Q\geq q^2$, so that
$f^{-1} \geq 0$. Note that $\omega$ is smooth at
finite $y$ since $\omega_\phi(x=\pm 1)=\omega_\psi(y=-1)=0$
(\ie there are no Dirac-Misner string pathologies). In
verifying that (\ref{simplef}, \ref{omegas}) solve \eqn{eom} it is
useful to observe that
$(1-x^2)\:\omega_{\psi , x} = (y^2-1)\: \omega_{\phi , y}$. The
Maxwell field strength $F=dA$ is uniquely determined by $f$ and $\omega$
\cite{harveyetal}. For our solution the gauge potential is
\beq
 A=\frac{\sqrt{3}}{2} \left[ f \, (dt+\omega) -
 \frac{q}{2} ((1+x) \, d\phi + (1+y) \, d\psi) \right] \,.
 \label{apot}
\eeq
The metric \reef{base} can be brought to a
manifestly flat form with the coordinate transformation
\be
\label{eqn:flatcoords}
\rho \sin \Theta = \frac{R \sqrt{y^2-1}}{x-y}, \qquad \rho \cos \Theta
= \frac{R  \sqrt{1-x^2}}{x-y}  \,.
\ee
In these coordinates the solution takes the
form
\beq
f^{-1}=1+\frac{Q-q^2}{\Sigma}+\frac{q^2 \rho^2}{\Sigma^2}
\label{fother}\,,
\eeq
\beqa
\omega_\phi &=& -\frac{q
\rho^2 \cos^2 \Theta}{2\Sigma^2}\left[3Q-q^2\left(
3-\frac{2 \rho^2}{\Sigma}\right)\right]\,,\nonumber\\
\omega_\psi &=&
-\frac{6qR^2 \rho^2 \sin^2 \Theta}{\Sigma(\rho^2+R^2+\Sigma)}\nonumber\\
&&-\frac{q
\rho^2 \sin^2 \Theta}{2\Sigma^2}\left[3Q-q^2\left(
3-\frac{2\rho^2}{\Sigma}\right)\right]\,,
\label{omegasother}\eeqa
where $\Sigma\equiv \sqrt{(\rho^2 - R^2)^2+4R^2 \rho^2 \cos^2
\Theta}$. Using these
coordinates it is straightforward to see that if we set $R= 0$ then 
the solution reduces to the BMPV black hole.

\noindent
{\bf Symmetries and charges.}
The results of \cite{harveyetal} imply that our black ring preserves four
supersymmetries. It has isometry group $\bR \times U(1)^2$, whereas BMPV
has $\bR \times U(1) \times SU(2)$.  The mass and angular momenta of the solution follow from its
manifestly asymptotically flat form (\ref{fother}, \ref{omegasother}) as
\beqa
M&=& \frac{3\pi}{4G}Q\,,\qquad
J_\phi=\frac{\pi}{8G} \, q \, (3Q-q^2) \, , \nonumber \\
J_\psi&=&\frac{\pi}{8G}\, q \, (6R^2+3Q-q^2)\, .
\label{jpsi}
\label{adm}
\eeqa
The total electric charge ${\bf Q}$ is proportional to $Q$ and
satisfies $M = (\sqrt{3}/2)
{\bf Q}$, hence the BPS inequality of \cite{gibbons:94} is saturated.

\noindent
{\bf Absence of closed timelike curves.}
As $y \rightarrow -\infty$ we find
\bea
&& g_{\psi\psi}= 3\left[\frac{(Q-q^2)^2}{4 q^2}-R^2 \right]
+\frac{q^2}{4}(1-x^2)+ {\cal O}\left( y^{-1} \right)\,, \nonumber
\eea
so we demand
\beq
R < \frac{Q-q^2}{2 q}
\label{noctcs}
\eeq
to ensure that $\partial/\partial \psi$ remains spacelike. This
condition is sufficient to avoid any CTCs at finite $y$. To see this,
consider the function $(x-y)^2 f^{-2} g_{\psi\psi}/(-1-y)$, which is a
polynomial in $x$ and $y$. This can be grouped into a sum of terms that
are all non-negative if \reef{noctcs} holds. In the same manner, one can
check that the determinant of the $2\times 2$ metric $g_{ij}$,
$i,j=\phi,\psi$ is always non-negative. These two conditions are
necessary and sufficient for $g_{ij}$ to be positive semi-definite.

\noindent
{\bf The event horizon.}
To examine what happens as $y \rightarrow - \infty$ it is
convenient to define a new coordinate $r = -R/y$. Now consider a
coordinate transformation of the form
\bea
&& dt = dv - B(r) dr, \qquad d\phi = d\phi' - C(r) dr, \nonumber \\
&& d\psi = d\psi' - C(r) dr,
\eea
where
\be
B(r) = \frac{B_2}{r^2} + \frac{B_1}{r} + B_0, \qquad C(r) =
\frac{C_1}{r} + C_0.
\ee
The electromagnetic potential is regular in the new coordinates up
to terms that can be removed by a gauge transformation. The
constants $B_i$ and $C_i$ will be chosen so that all metric
components remain finite as $r \rightarrow 0$. To eliminate a $1/r$
divergence in $g_{r\psi'}$ and a $1/r^2$ divergence in $g_{rr}$ we
choose $B_2 = q^2L/(4R)$ and $C_1 = -q/(2L)$, where
\be
L \equiv \sqrt{3 \left[ \frac{(Q-q^2)^2}{4q^2} - R^2 \right] },
\ee
which is positive as a consequence of (\ref{noctcs}). To avoid a
$1/r$ divergence in $g_{rr}$ we need $ B_1 = (Q+2q^2)/(4 L) + L
(Q-q^2)/(3 R^2)$. The metric is then analytic at $r=0$ with $g_{rr}$
a linear function of $x$ at $r=0$. We can eliminate this function by
choosing the finite part of the coordinate transformation as
follows: $C_0 = -(Q-q^2)^3/(8q^3RL^3)$, $B_0 = q^2 L / (8 R^3) +
2L/(3R) -R/(2L) + 3 R^3/(2L^3) + 3(Q-q^2)^3/(16q^2RL^3)$. The metric
can now be written
\bea
&& ds^2 = -\frac{16r^4}{q^4} dv^2 +
\frac{2R}{L}  dv dr +  \frac{4 r^3 \sin^2 \theta}{Rq} dv d\phi'
\nonumber \\  &+& \frac{4R r}{q}  dv d\psi' 
+ \frac{3qr \sin^2 \theta}{L} drd\phi' 
\nonumber \\ 
&+& 2\Big[ \frac{qL}{2R} \cos\theta
+ \frac{3 q R}{2L} +  \frac{(Q-q^2)(3R^2-2L^2)}{3qRL}
 \Big] dr d\psi' \nonumber \\
&+& L^2 d{\psi'}^2 + \frac{q^2}{4}
\left[d\theta ^2 + \sin^2 \theta \left( d\phi' - d\psi' \right)^2
\right] + \ldots
\eea
where $x = \cos \theta$ and 
the ellipsis denotes terms involving 
subleading (integer) powers of $r$ in all of the metric components explicitly
indicated, as well as terms in $g_{rr}$ starting at ${\cal O}(r)$. (We have not displayed the leading order term in $g_{rr}$ because it is lengthy and unilluminating.)
The determinant of this metric is analytic in $r$. At $r=0$ it
vanishes if, and only if, $\sin^2 \theta = 1$, which is just a
coordinate singularity. It follows that the inverse metric is also 
analytic in $r$ and hence the above coordinates define an analytic 
extension of our solution through the surface $r=0$.

The supersymmetric Killing vector field $V =
\partial/\partial v$ is null at $r=0$. Furthermore
$V_\mu dx^\mu = (R/L) dr$ at $r=0$, so $V$ is normal to the surface
$r=0$. Hence $r=0$ is a null hypersurface and a Killing horizon of
$V$, i.e., the black ring has an event horizon at $r=0$.

If $L=0$ then a similar analysis shows that the geometry has a
null orbifold singularity instead of an event horizon.
\newline
\noindent {\bf Horizon geometry.} We can read off the geometry of a
spatial cross-section of the event horizon: 
\beq 
ds^2_\rom{horizon} =
L^2 d{\psi'}^2 +\frac{q^2}{4} \left(d\theta^2 + \sin^2{\theta} d\chi^2
\right) \, , 
\eeq 
where $\chi \equiv \phi' - \psi' = \phi - \psi$. We see that
the horizon has geometry $S^1 \times S^2$, where the $S^1$ and $S^2$ have
radii $L$ and $q/2$, respectively. Note that the $S^2$ is round, in
contrast with non-extremal black rings, for which the $S^2$ is deformed
in the $\theta$ direction. 

The area of the event horizon is
\beq
{\cal A}= 2\pi^2 L q^2 = \pi^2 q
 \sqrt{ 3 \left[ (Q-q^2)^2-4 q^2 R^2\right]}\,.
\eeq
The surface gravity and angular velocities of the event horizon vanish,
as expected for a supersymmetric, asymptotically flat black hole
\cite{gauntlett:99}. The horizon is at infinite proper spatial distance
from points outside it, \ie it lies down an infinite throat.
\newline
\noindent
{\bf Near-horizon limit.}
The near-horizon limit is defined by $r = \e L \tilde{r}/R$, $v =
\tilde{v}/\e$ and $\e \rightarrow 0$. In this limit, the metric
becomes
\ba
ds^2 &=& 2 d\tilde{v} d\tilde{r} + \frac{4L}{q} \tilde{r}
d\tilde{v} d\psi' + L^2 d{\psi'}^2 \nonumber \\
&+& \frac{q^2}{4} \left( d\theta^2 +
\sin^2 \theta d\chi^2 \right).
\ea
This is the product of a spacetime that is locally $AdS_3$ with radius
$q$ and a 2-sphere of radius $q/2$, as expected from \cite{reall:02}.
The $AdS_3$ space is the 
near-horizon geometry of an extremal BTZ black hole of horizon radius
$r_+ = L$. This near-horizon limit is not the
same as the `decoupling limit' relevant to the
AdS$_3$/CFT$_2$ duality, which will be analyzed in \cite{EEMR}.


\noindent
{\bf Infinite-radius limit.} The existence of a supersymmetric black
ring solution was recently conjectured in \cite{benakraus,bena}. As
evidence, \cite{bena} constructed a black string solution that was claimed to
describe a black ring of infinite radius. This solution can indeed
be recovered as the infinite radius limit of our solution as
follows.  Define a charge density $\bar{Q} = Q/2R$
and new coordinates $r=-R/y$, $\eta=R\psi$, $\cos\theta=x$. In the
limit $R \rightarrow \infty$ with $\bar{Q}$, $q$, $r$, and $\eta$
held fixed we have
\beqa
f^{-1} &\to& 1+ \frac{\bar{Q}}{r} + \frac{ q^2}{4 r^2} \, ,\\[2mm]
\omega_\psi d\psi &\to&
 -\left(  \frac{3 q}{2 r}
  + \frac{3 q \bar{Q}}{4 r^2}
  + \frac{ q^3}{8 r^3} \right) d\eta \, ,
\eeqa
and $\omega_\phi \to 0$.
This is the solution of \cite{bena} for the special case of three
equal charges and three equal dipoles.
In \cite{EEMR} we shall present a more general supersymmetric black
ring whose infinite-radius limit is the general solution of
\cite{bena}. Note that the solution of \cite{bena} does not
exhibit an essential feature of the supersymmetric black ring, namely
the existence of {\it two} independent angular momenta.

\begin{figure}[thb]
\begin{picture}(0,0)(0,0)
\put(4,-6){0}
\put(4,131){$a_H$}
\put(-9,60){$j_\phi$}
\put(105,-9){$j_\psi$}
\put(-16,105){$\frac{1}{2\sqrt{2}}$}
\end{picture}
\centering{\psfig{file=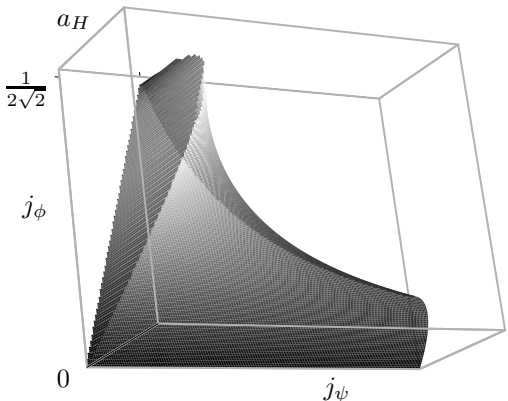,width=6cm}}
\vspace{3ex}
\caption{Plot of the dimensionless horizon area $a_H={\cal A}/(GM)^{3/2}$ as
a function of the dimensionless angular
momenta $j_i =(27 \pi /(32 G))^{1/2} J_i/M^{3/2}$, $i = \psi,\phi$. The
scales for $j_\phi$ and $j_\psi$ are different for a better
representation: the planar boundary corresponds to $j_\psi=j_\phi$
(which is only reached as $R \to 0$). The surface extends to infinity to
the right.}
\label{horizon}
\end{figure}

\noindent
{\bf Dipole charge.} Define
\be
{\cal D} = \frac{1}{16\pi G} \int_{S^2} F = \frac{\sqrt{3}}{16 G} q \, ,
\ee
where the $S^2$ is a surface of constant
$t$, $\psi$ and $y$ outside the horizon. This `charge' determines the
radius of the $S^2$ of the horizon and also (for fixed electric charge)
the angular momentum $J_{\phi}$. It is not conserved except in the limit
in which the ring becomes an infinite black string. 
The general solution of \cite{EEMR} carries three independent dipole
charges, which are proportional respectively to the number of
D1-branes, D5-branes
and Kaluza-Klein (KK) monopoles with a worldvolume direction around the
ring circle. The solution presented here corresponds to taking equal
values for these three dipole charges, so $q$ is proportional to
the number of branes with a worldvolume direction around the ring
circle. When oxidized to six dimensions, the black ring becomes a black
supertube. We anticipate that regularity of this solution
will lead to $q$ being quantized in units of the radius of
the KK circle, since $q$ is the number of KK monopoles making up the
tube \cite{EEMR}.
\newline
\noindent
{\bf Uniqueness.} The supersymmetric black ring is uniquely specified by
its electric charge and angular momenta. Figure \ref{horizon} shows the
region of the $J_\phi$-$J_\psi$ plane occupied by BPS black rings for
fixed charge $Q$. There are three boundaries to this region. 
The boundary to the upper right arises from the condition \reef{noctcs} 
with $L$, the radius of the $S^1$, vanishing at the boundary. The
lower boundary $J_{\phi} \rightarrow 0$ arises from the condition
$q>0$ so the radius of the $S^2$ vanishes at this boundary. The leftmost,
straight boundary arises from the condition $R>0$, which implies $J_\psi
> J_\phi$. If $R=0$ the solution reduces to the BMPV solution, so there
are no black rings with $J_\psi = J_\phi$, and thus the conserved
charges of our black ring are always different from those
of a BMPV black hole.

\noindent {\bf Entropy.}
Figure \ref{horizon} displays the black ring horizon area as a
function of the angular momenta for fixed charge $Q$. For fixed
$J_\phi$ the entropy function is maximized as $R \rightarrow 0$.
However, this function is not continuous at $R=0$. At this point the
black ring solution reduces to the BMPV black hole, whose entropy is
greater than the $R \rightarrow 0$ limit of the black ring entropy.
This discontinuity is due to the change in the horizon topology from
$S^1 \times S^2$ at $R>0$ to $S^3$ at $R=0$, and is analogous to the
discontinuous increase in entropy that occurs when two sources of a
multi-centre extremal Reissner-Nordstrom solution become coincident.

The existence of supersymmetric black rings raises many questions. For
example: Are there any supersymmetric black hole solutions 
(of minimal supergravity)
for the empty regions of the $J_\phi$-$J_\psi$ plane, or any
solutions that overlap with the currently covered regions? If not,
then can the results of \cite{reall:02} be strengthened to a
full uniqueness theorem for supersymmetric black holes?
Is there a general non-extremal 
black ring solution that reproduces our solution and those of
\cite{ER,HE} as special cases? Such a solution would presumably
depend on 5 parameters corresponding to the angular momenta, the
electric charge, the dipole charge and the mass. Finally: Is
it possible to perform a statistical-mechanical calculation of the
entropy of this black ring?

\medskip
\noindent
We thank V. Balasubramanian, B.\ Cabrera-Palmer, J.\ Gauntlett, 
D.\ Marolf, J. Sim\'on, and especially G.\ Horowitz and R. Myers, for
useful discussions. HE was supported by the Danish Research Agency
and NSF grant PHY-0070895. RE was supported in part by
UPV00172.310-14497, FPA2001-3598, DURSI 2001-SGR-00188,
HPRN-CT-2000-00131. HSR was supported in part by the National
Science Foundation under Grant No.~PHY99-07949.



\begin{thebibliography}{99}

\bibitem{stva}
A.~Strominger and C.~Vafa,
Phys.\ Lett.\ B {\bf 379}, 99 (1996)
[arXiv:hep-th/9601029].

\bibitem{bmpv}
J.~C.~Breckenridge, R.~C.~Myers, A.~W.~Peet and C.~Vafa,
Phys.\ Lett.\ B {\bf 391}, 93 (1997)
[arXiv:hep-th/9602065].

\bibitem{ER}
R.~Emparan and H.~S.~Reall,
Phys.\ Rev.\ Lett.\ {\bf 88}, 101101 (2002) [arXiv:hep-th/0110260].

\bibitem{reall:02}
H.~S.~Reall,
Phys.\ Rev.\ D {\bf 68}, 024024 (2003), erratum {\bf 70}, 089902 (2004) [arXiv:hep-th/0211290]. 

\bibitem{benakraus}
I.~Bena and P.~Kraus,
Phys.\ Rev.\ D {\bf 70}, 046003 (2004)
[arXiv:hep-th/0402144].

\bibitem{mathur}
O.~Lunin and S.~D.~Mathur,
Phys.\ Rev.\ Lett.\  {\bf 88}, 211303 (2002)
[arXiv:hep-th/0202072],
S.~D.~Mathur, A.~Saxena and Y.~K.~Srivastava,
Nucl.\ Phys.\ B {\bf 680}, 415 (2004)
[arXiv:hep-th/0311092].

\bibitem{MT}
D.~Mateos and P.~K.~Townsend,
Phys.\ Rev.\ Lett.\ {\bf 87} (2001) 011602 [arXiv:hep-th/0103030].

\bibitem{EEMR}
H.~Elvang, R.~Emparan, D.~Mateos and H.~S.~Reall, 
hep-th/0408120.

\bibitem{HE}
H.~Elvang,
Phys.\ Rev.\ D {\bf 68}, 124016 (2003) [arXiv:hep-th/0305247].
%
H.~Elvang and R.~Emparan,
JHEP {\bf 0311}, 035 (2003) [arXiv:hep-th/0310008].
%
R.~Emparan,
JHEP {\bf 0403}, 064 (2004) [arXiv:hep-th/0402149].


\bibitem{gibbons:94}
G.~W.~Gibbons, D.~Kastor, L.~A.~J.~London, P.~K.~Townsend and J.~H.~Traschen,
Nucl.\ Phys.\ B {\bf 416}, 850 (1994)
[arXiv:hep-th/9310118].

\bibitem{harveyetal}
J.~P.~Gauntlett, J.~B.~Gutowski, C.~M.~Hull, S.~Pakis and H.~S.~Reall,
Class.\ Quant.\ Grav.\ {\bf 20}, 4587 (2003)
[arXiv:hep-th/0209114]. We follow the conventions
of this paper except for the metric signature, which we take to be
positive.

\bibitem{gauntlett:99}
J.~P.~Gauntlett, R.~C.~Myers and P.~K.~Townsend,
Class.\ Quant.\ Grav.\ {\bf 16}, 1 (1999) [arXiv:hep-th/9810204].

\bibitem{bena}
I.~Bena,
arXiv:hep-th/0404073.

\end{thebibliography}
\end{document}